\documentclass[aps,prb, twocolumn] {revtex4}
\usepackage{epsfig}

\begin{document}
\headsep 2.5cm
\title{Ferromagnetic semiconductor single wall carbon nanotube.}

\author{Rostam Moradian$^{1,2}$}
\email{rmoradian@razi.ac.ir}
 \author{Ali Fathalian$^{1}$}

\affiliation{$^{1}$Physics Department, Faculty of Science, Razi
University, Kermanshah, Iran\\
$^{2}$Computational Physical Science Research Laboratory, Department of Nano-Science, Institute for Studies in Theoretical Physics and Mathematics (IPM)
            ,P.O.Box 19395-1795, Tehran, Iran}

\date{\today}

\begin{abstract}
Possibility of a ferromagnetic semiconductor single wall carbon nanotube (SWCNT) , where ferromagnetism is due to coupling between doped magnetic impurity on a zigzag SWCNT and electrons spin, is investigate. We found, in the weak impurity-spin couplings, at low impurity concentrations the spin up electrons density of states remain semiconductor while the spin down electrons density of states shows a metallic behavior. By increasing impurity concentrations the semiconducting gap of spin up electrons in the density of states is closed, hence a semiconductor to metallic phase transition is take place. In contrast, for the case of strong coupling, spin up electrons density of states remain semiconductor and spin down electron has metallic behavior. Also by increasing impurity spin magnitude, the semiconducting gap of spin up electrons is increased.
\end{abstract}


\maketitle
\section {Introduction}
Single wall carbon nanotubes extensively attracted from both theoretical and experimental point of view, due to it's technological applications such as nano electronics devises\cite{Paul:00, Tans:98}, nano ring (SWCNT ring)\cite{Cuniberti:02}, molecule-nanotube hybrids\cite{Guntierrez:03}, quantum dots\cite{Tans:97}. Effect of substitution non magnetic impurity on a SWCNT are investigated by many peoples \cite{Harigaya:99, Latil:04, MoradianBNC:04}. We found that by doping boron and nitrogen the semiconductor gap, $E_{g}$, of a zigzag SWCNT could be control\cite{MoradianBNC:04}. It is shown that vacancy could lead to electron spin polarization, hence a magnetic SWCNT\cite{Markarova:01, Coey:02}. But role of doping finite magnetic impurity concentration on a zigzag SWCNT is not challenged.  Our motivation is to investigate possibility of a ferromagnetic semiconductor SWCNT in comparison to the dilute ferromagnetic semiconductors\cite{Ohno:98}.
This paper is organized as follows: In Sec. II model and treatment formalism of a magnetic impurity doped zigzag SWCNT is introduced. In Sec. III three cases are considered: in the first case just impurity concentration is changed while impurity spin magnitude, exchange coupling and temperature are fixed. We found for the weak exchange coupling, $J$, by increasing impurity concentration the semiconductor gap, $E^{\uparrow}_{g}$, in the spin up density of states is closed hence a semiconductor to metal phase transition is happened. In the second case just exchange coupling is varied and other parameters are fixed, we found by increasing exchange coupling ,J, the semiconducting gap, $E^{\uparrow}_{g}$, of spin up electrons in their density of states is increased. In the last case just magnitude of spin, $S$, is changed. In this case we found by increasing $S$ the semiconducting gap, $E^{\uparrow}_{g}$, in the spin up density of states is increased.     

\section{Model and formalism}
Let us consider the Hamiltonian as a general random tight-binding model\cite{MoradianBNC:04}        
\begin{eqnarray}
H=-\sum_{ij\alpha\beta\sigma^{'}}t^{\alpha\beta}_{i\sigma j\sigma^{'}}{c^{\alpha}}^{\dagger}_{i\sigma}c^{\beta}_{j\sigma^{'}}+\sum_{i\alpha\sigma} (\varepsilon^{\alpha}_{i}-\mu)
 \hat{n}^{\alpha}_{i\sigma}\nonumber\\+\sum_{i\alpha}J_{i}{\bf S}^{\alpha}_{i}.{\bf s}^{\alpha}_{i}+\sum_{i\alpha\sigma} u_{i}
 \hat{n}^{\alpha}_{i\sigma}\hat{n}^{\alpha}_{i-\sigma},
\label{eq:Hamiltonian}
\end{eqnarray}
where $\alpha$ and $\beta$ refer to the A or B sites inside of the graphene Bravais lattice unit cell, where each Bravais lattice site is includes two nonequivalent sites that are indicated by A and B, ${c^{\alpha}}^{\dagger}_{i\sigma}$ ($c^{\alpha}_{i\sigma}$) is the creation (annihilation) operator of an electron with spin $\sigma$ on Bravais lattice site $i$, and $\hat{n}^{\alpha}_{i\sigma}={c^{\alpha}}^{\dagger}_{i\sigma}c^{\alpha}_{i\sigma}$ is the number operator. $t^{\alpha\beta}_{i\sigma j\sigma^{'}}$ are the hopping integrals between the $\pi$ orbitals of sites $i$ and $j$ with spin $\sigma$ and $\sigma^{'}$ respectively. $\mu$ is the chemical potential, $\varepsilon^{\alpha}_{i}$ is the random on-site energy where it takes  $0$ with probability $1-c$ for host sites and $\delta$ with probability $c$ for impurity sites. $J_{i}$ is random exchange coupling between $\pi$ electrons and impurity spins where it takes  $0$ with probability $1-c$ for host sites and $J$ with probability $c$ for impurity sites. ${\bf S}^{\alpha}_{i}$ is the impurity spin operator and ${\bf s}^{\alpha}_{i}=\sum_{\sigma\sigma^{'}}{c^{\alpha}}^{\dagger}_{i\sigma}{\mbox{\boldmath $\tau$}}_{\sigma\sigma^{'}}{c^{\alpha}}_{i\sigma^{'}}$ is the electron spin operator. Where ${\mbox{\boldmath $\tau$}}$ is Pauli matrix vector and $u_{i}$ is the Coulomb interaction between electrons with opposite spin. The matrix form of Eq.\ref{eq:Hamiltonian} is, 
\begin{equation}
H=-\sum_{ij}{\Psi}^{\dagger}_{i}{\bf t}_{ij}\Psi_{j}+\sum_{i} {\Psi}^{\dagger}_{i} ({\bf V}_{i}-\mu{\bf I}){\Psi}_{i}+\sum_{i} {\Psi}^{\dagger}_{i}  u_{i}{\Psi}_{i},
\label{eq:matrixHamiltonian}
\end{equation}
where the two-component field operator, ${\Psi}^{\dagger}_{i\sigma}$, is given by
\begin{equation}
\Psi_{i}= \left(
       \begin{array}{c}
c^{A}_{i \uparrow} \\c^{B}_{i \uparrow} \\c^{A}_{i \downarrow} \\c^{B}_{i \downarrow}
\end{array}\right),
\label{eq:fourtwosite field}
\end{equation}
 ${\bf V}_{i}$ is the random on-site energy matrix operator,
\begin{eqnarray}
{\bf V}_{i}&=&\small{\left(
       \begin{array}{cccc}
\varepsilon^{A}_{i}+u_{i}n^{A}_{i\uparrow} & 0 & 0  & 0\\
0 &\varepsilon^{B}_{i}+u_{i}n^{B}_{i\uparrow} & 0 &  0\\
 0 & 0 & \varepsilon^{A}_{i}+u_{i}n^{A}_{i\downarrow}  & 0 \\
0 & 0 & 0 &\varepsilon^{B}_{i}+u_{i}n^{B}_{i\downarrow}\\
\end{array}\right)}\nonumber\\
&+&
\small{\left(
       \begin{array}{cccc}
\frac{1}{2}J^{A}_{i}S^{Az}_{i} & 0 & \frac{1}{2}J^{A}_{i}S^{A-}_{i} & 0\\
0 &\frac{1}{2}J^{B}_{i}S^{B z}_{i} & 0 &  \frac{1}{2}J^{B}_{i}S^{B-}_{i}\\
 \frac{1}{2}J^{A}_{i}S^{A+}_{i}& 0 & -\frac{1}{2}J^{A}_{i}S^{Az}_{i}  & 0 \\
0 & \frac{1}{2}J^{B}_{i}S^{B+}_{i} & 0 &-\frac{1}{2}J^{B}_{i}S^{Bz}_{i}\\
\end{array}\right)},
\label{eq:fourtwosite random energy}
\end{eqnarray}
 and ${\bf{t}}_{ij}$  is the hopping matrix in the spinor space defined by,
\begin{equation}
{\bf{t}}_{ij}=\left(
       \begin{array}{cccc}
t^{AA}_{i\uparrow j \uparrow} & t^{AB}_{i \uparrow j \uparrow} & t^{AA}_{i \uparrow j \downarrow} & t^{AB}_{i \uparrow j \downarrow}\\
 t^{BA}_{i\uparrow j\uparrow}&t^{BB}_{i\uparrow j\uparrow} &  t^{BA}_{i\uparrow j\downarrow }&t^{BB}_{i\uparrow j\downarrow}\\
t^{AA}_{i\downarrow j\uparrow} & t^{AB}_{i\downarrow j\uparrow} & t^{AA}_{i\downarrow j\downarrow} & t^{AB}_{i\downarrow j\downarrow}\\
 t^{BA}_{i\downarrow j\uparrow}&t^{BB}_{i\downarrow j\uparrow} &  t^{BA}_{i\downarrow j\downarrow}&t^{BB}_{i\downarrow j\downarrow}\\

\end{array}\right),
\label{eq:fourtwosite hopping integral}
\end{equation}
 and ${\bf I}$ is a $4\times 4$ unitary matrix.

 The electron's equation of motion corresponding to Eq.\ref{eq:Hamiltonian} in the mean field approximation where the operator ${\bf S}_{i}$ is replaced by it's thermal averaged, ${\bf S}_{i}\approx \langle {\bf S}_{i}\rangle_{th}$, is
\begin{equation}
\sum_{l} \left(
       \begin{array}{c}
(E{\bf I}-{\mbox{\boldmath $\varepsilon$}}_{i}+{\bf I}\mu_{i})\delta_{il}-{\bf t}_{il}\end{array}\right){\bf G}(l,j;E)={\bf I}\delta_{ij}
\label{eq:equation of motion}
\end{equation}
where ${\bf G}(i,j; E)$ is the  random Green function matrix defined by,
\begin{equation}
 {\bf G}(i,j; E)=\left(
       \begin{array}{cccc}
G^{AA}_{i\uparrow j \uparrow} & G^{AB}_{i \uparrow j \uparrow} & G^{AA}_{i \uparrow j \downarrow} & G^{AB}_{i \uparrow j \downarrow}\\
 G^{BA}_{i\uparrow j\uparrow}& G^{BB}_{i\uparrow j\uparrow} &  G^{BA}_{i\uparrow j\downarrow }& G^{BB}_{i\uparrow j\downarrow}\\
t^{AA}_{i\downarrow j\uparrow} & G^{AB}_{i\downarrow j\uparrow} & G^{AA}_{i\downarrow j\downarrow} & G^{AB}_{i\downarrow j\downarrow}\\
 G^{BA}_{i\downarrow j\uparrow}& G^{BB}_{i\downarrow j\uparrow} &  G^{BA}_{i\downarrow j\downarrow}& G^{BB}_{i\downarrow j\downarrow}\\
\end{array}\right).
\label{eq:fourtwosite Green function}
\end{equation}
Here after spin flip is neglected, hence Eq.\ref{eq:fourtwosite Green function} is reduced to the two following separated equations for spin up and down electrons,
\begin{equation}
\sum_{l} \left(
       \begin{array}{c}
(E{\bf I}-{\mbox{\boldmath $\varepsilon$}}_{i\sigma}+\mu_{i}{\bf I})\delta_{il}-{\bf t}_{i\sigma l\sigma}\end{array}\right){\bf G}_{\sigma\sigma}(l,j;E)={\bf I}\delta_{ij}
\label{eq:twoequation of motion}
\end{equation}
where ${\bf G}_{\sigma\sigma}(i,j; E)$ is a $2\times 2$ random Green function matrix for spin $\sigma$ defined by,
\begin{equation}
 {\bf G}_{\sigma\sigma}(i,j; E)=\left(
       \begin{array}{cc}
 G^{AA}_{\sigma\sigma}(i,j; E) & G^{AB}_{\sigma\sigma}(i,j; E)\\
 G^{BA}_{\sigma\sigma}(i,j; E) & G^{BB}_{\sigma\sigma}(i,j; E)
\end{array}\right),
\label{eq:twosite Green function}
\end{equation}
${\bf t}_{i\sigma l\sigma}$ is hoping integral matrix for spin $\sigma$ which is given by,
\begin{equation}
 {\bf t}_{i\sigma j\sigma}=\left(
       \begin{array}{cc}
 t^{AA}_{i\sigma j\sigma} & t^{AB}_{i\sigma j\sigma}\\
 t^{BA}_{i\sigma j\sigma} & t^{BB}_{i\sigma j\sigma}
\end{array}\right),
\label{eq:twosite hopping}
\end{equation}
 and ${\mbox{\boldmath $\varepsilon$}}_{i\sigma}$ is random on site potential matrix for spin $\sigma$,
\begin{eqnarray}
{\mbox{\boldmath $\varepsilon$}}_{i\sigma}&=&\left(
       \begin{array}{cc}
\varepsilon^{A}_{i}+u_{i}n^{A}_{i\sigma} & 0 \\
0 &\varepsilon^{B}_{i}+u_{i}n^{B}_{i\sigma}  \\
\end{array}\right)
 \nonumber\\  &+&   
\left( \begin{array}{cc}
\frac{1}{2}q_{\sigma}J^{A}_{i}\langle S^{Az}_{i}\rangle_{th} & 0 \\
0 & \frac{1}{2}q_{\sigma}J^{B}_{i}\langle S^{B z}_{i}\rangle_{th}  \\
\end{array}\right)
\label{eq:twosite random energy}
\end{eqnarray}
where $q_{\sigma}=1$ for spin up and -1 for spin down electrons.
Eq.\ref{eq:twoequation of motion}, could be rewritten in terms of the perfect Green function matrix ${\bf G}^{0}(i,j;E)$ as,
\begin{equation}
 {\bf G}_{\sigma\sigma}(i,j;E)={ \bf G}^{0}(i,j;E)+\sum_{l}{\bf G}^{0}(i,l;E)
{\mbox{\boldmath $\varepsilon$}}_{l\sigma}{\bf G}_{\sigma\sigma}(l,j;E)
\label{eq:expanding interms of random onsite potential}
\end{equation}
where ${\bf G}^{0}(i,j;E)$ is given by

\begin{equation}
{\bf G}^{0}(i,j;E)=\frac{1}{N}\sum_{\bf k}e^{\imath{\bf k}.{\bf r}_{ij}}
\left(\begin{array}{c}
E{\bf I}-{\mbox{\boldmath $\epsilon$}}_{\bf k}
\end{array}\right)^{-1}.
\label{eq:clean}
\end{equation}
with ${\mbox{\boldmath $\epsilon$}}_{\bf k}=-\mu{\bf I}+\frac{1}{N}\sum_{ij}{\bf t}_{i\sigma j\sigma}e^{\imath{\bf k}.{\bf r}_{ij}}$ is the band structure for perfect system. In our calculations the hopping randomness are neglected also we allowed  hopping to the nearest neighbors and neglected others, so
\begin{equation}
t^{AB}_{<i\sigma j\sigma>}= t^{BA}_{<i\sigma j\sigma>}=t,
\label{eq:tclean}
\end{equation}
where $t\sim 2.5eV$ is clean system nearest neighbour hopping integral. Hence 
\begin{equation}
{\bf t}_{<i\sigma j\sigma>}=\left(
       \begin{array}{cc}
0 & t^{AB}_{<i\sigma j\sigma>} \\
 t^{BA}_{<i\sigma j\sigma>}& 0
\end{array}\right),
\label{eq:twosite hopping}
\end{equation}
and the dispersion relation is
\begin{equation}
{\mbox{\boldmath $\epsilon$}}_{\bf k}=\left(
       \begin{array}{cc}
-\mu & t\gamma({\bf k}) \\
 t\gamma^{*}({\bf k}) & -\mu
\end{array}\right).
\label{eq:dispersion relation}
\end{equation}
where $\gamma({\bf k})=\sum_{i=1}^{3}e^{\imath{\bf k}.{\bf r}_{i}}$ and ${\bf r}_{i}$ are three vectors that connect an A (B) site to it's nearest neighbors B(A) sites.

The Dyson equation for the averaged Green function, $\bar{\bf G}_{\sigma\sigma}(i,j;E)$, corresponding to Eq.\ref{eq:expanding interms of random onsite potential} is  
\begin{eqnarray}
 \bar{\bf G}_{\sigma\sigma}(i,j;E)&=&{\bf G}^{0}(i,j;E)\nonumber \\
&+&\sum_{ll^{'}}{\bf G}^{0}(i,l;E)
{\bf\Sigma}_{\sigma\sigma}(l,l^{'};E){\bar{\bf G}}_{\sigma\sigma}(l^{'},j;E),\nonumber \\
\label{eq:Dyson equation}
\end{eqnarray}
where the self energy ${\bf\Sigma}_{\sigma\sigma}(l,l^{'};E)$ is defined by
\begin{equation}
\langle {\mbox{\boldmath $\varepsilon$}}_{l\sigma}{\bf G}_{\sigma\sigma}(l,j;E)\rangle = \sum_{l^{'}}{\bf\Sigma}_{\sigma\sigma}(l,l^{'};E){\bar{\bf G}}_{\sigma\sigma}(l^{'},j;E).
\label{eq:self energy definition}
\end{equation}
The Fourier transform of  $\bar{\bf G}_{\sigma\sigma}(i,j;E)$ in Eq.\ref{eq:Dyson equation} is given by, 
 \begin{equation}
{ \bar {\bf G}}_{\sigma\sigma}(i,j;E)=\frac{2}{N}\sum_{\bf k}e^{\imath{\bf k}.{\bf r}_{ij}}
\left(\begin{array}{c}
E {\bf I}-{\mbox{\boldmath $\epsilon$}}_{\bf k}-{\bf\Sigma}_{\sigma\sigma}({\bf k};E) 
\end{array}\right)^{-1} 
\label{eq:exact average green function}
\end{equation}
where 
 \begin{equation}
{\bf\Sigma}_{\sigma\sigma}({\bf k};E)=\frac{2}{N}\sum_{i,j}e^{-\imath{\bf k}.{\bf r}_{ij}}
{\bf\Sigma}_{\sigma\sigma}(i,j;E),
\label{eq:self energy fourier transform}
\end{equation}
is the self energy Fourier transform. Since Eqs.\ref{eq:exact average green function}, \ref{eq:Dyson equation} and \ref{eq:expanding interms of random onsite potential} could not be solved exactly, we solve these equations in the coherent potential approximation (CPA). In the CPA multiple scattering is neglected and all sites are replaces by effective sites except one which is denoted by {\em impurity}, hence self energy is diagonal ${\bf\Sigma}_{\sigma\sigma}(i,j; E)={\bf\Sigma}_{\sigma\sigma}(E)\delta_{ij}$.  So Eq.\ref{eq:exact average green function}, \ref{eq:expanding interms of random onsite potential} and \ref{eq:self energy definition} at impurity site are reduce to,
 \begin{equation}
{ \bar {\bf G}}_{\sigma\sigma}(i,i;E)=\frac{2}{N}\sum_{\bf k}
\left(\begin{array}{c}
E {\bf I}-{\mbox{\boldmath $\epsilon$}}_{\bf k}-{\bf\Sigma}_{\sigma\sigma}(E) 
\end{array}\right)^{-1} ,
\label{eq:cpaexact average green function}
\end{equation} 

 \begin{eqnarray}
{\bf G}^{imp}_{\sigma\sigma}(i,i;E)&=&{\bar {\bf G}}_{\sigma\sigma}(i,i;E)\nonumber\\
&+& {\bar {\bf G}}_{\sigma\sigma}(i,i;E)( {\mbox{\boldmath $\varepsilon$}}_{i\sigma}-{\bf\Sigma}_{\sigma\sigma}(E)){\bf G}^{imp}_{\sigma\sigma}(i,i;E)\nonumber\\
\label{eq:impuritycpa average green function}
\end{eqnarray} 
and
\begin{equation}
\langle {\mbox{\boldmath $\varepsilon$}}_{i\sigma}{\bf G}^{imp}_{\sigma\sigma}(i,i;E)\rangle = {\bf\Sigma}_{\sigma\sigma}(E){\bar{\bf G}}_{\sigma\sigma}(i,i;E).
\label{eq:cpaself energy definition}
\end{equation}
Eqs.\ref{eq:cpaexact average green function}, \ref{eq:impuritycpa average green function} and \ref{eq:cpaself energy definition} are construct a complete set of equations that should be solved self consistently to provide ${\bar{\bf G}}_{\sigma\sigma}(i,i;E)$ and ${\bf\Sigma}_{\sigma\sigma}(E)$. In our calculations the impurity spin thermal average, $\langle S^{\alpha z}_{i}\rangle_{th}$, is evaluated quantum mechanically as,
\begin{equation}
\langle S^{\alpha z}_{i}\rangle_{th} =\frac{\sum^{S^{\alpha z}_{i}=S}_{S^{\alpha z}_{i}=-S}S^{\alpha z}_{i} exp({-\frac{1}{2}J\beta(n^{\alpha}_{i\uparrow}-n^{\alpha}_{i\downarrow})} S^{\alpha z}_{i})}{\sum^{S^{\alpha z}_{i}=S}_{S^{\alpha z}_{i}=-S} exp({-\frac{1}{2}J\beta(n^{\alpha}_{i\uparrow}-n^{\alpha}_{i\downarrow})} S^{\alpha z}_{i})},
\label{eq:spin thermal average}
\end{equation}
and the average magnetization is calculated from,
\begin{equation}
{\bar M}=\sum_{\xi} P_{\xi} \;\langle S^{\alpha z}_{i}\rangle_{th-\xi}
\label{eq:average magnetization}
\end{equation}
where $\xi$ take all $2^{2}$ impurity configurations and $P_{\xi}$ is probability of $\xi$ configuration.
\section{results and discussion}
\begin{figure}
\centerline{\epsfig{file=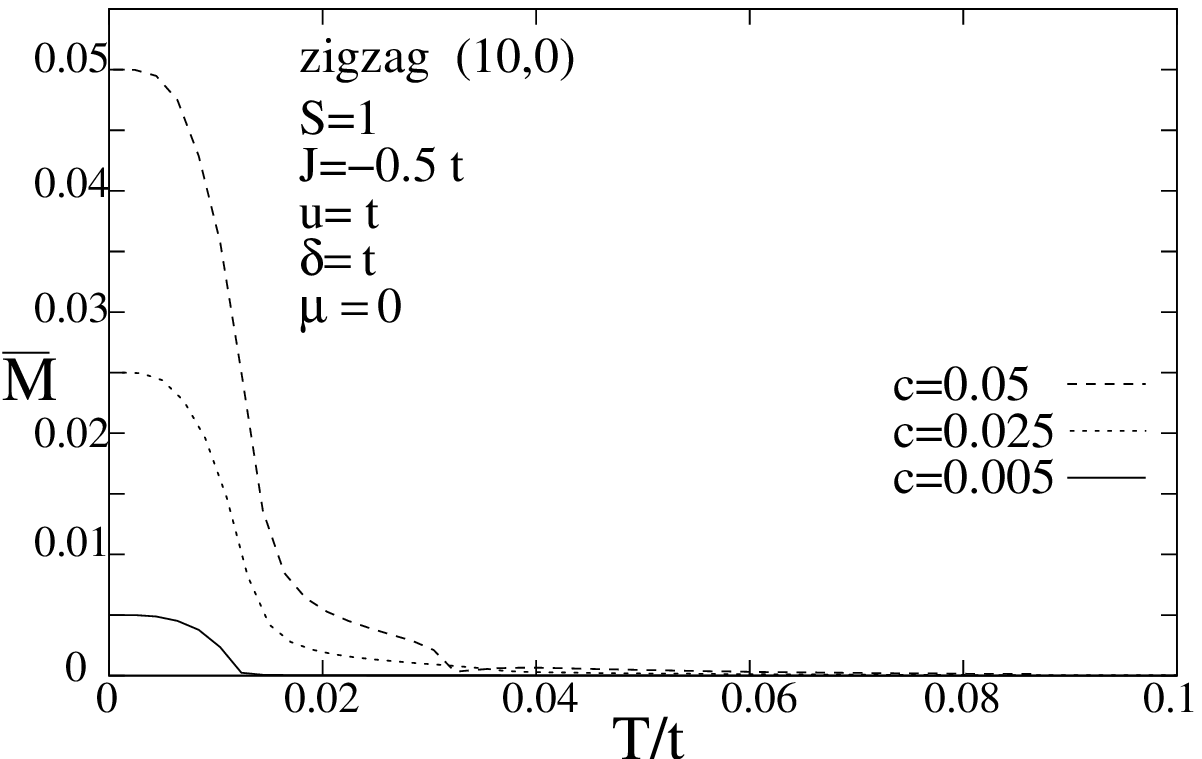 ,width=7.5cm,angle=0}}
\centerline{\epsfig{file=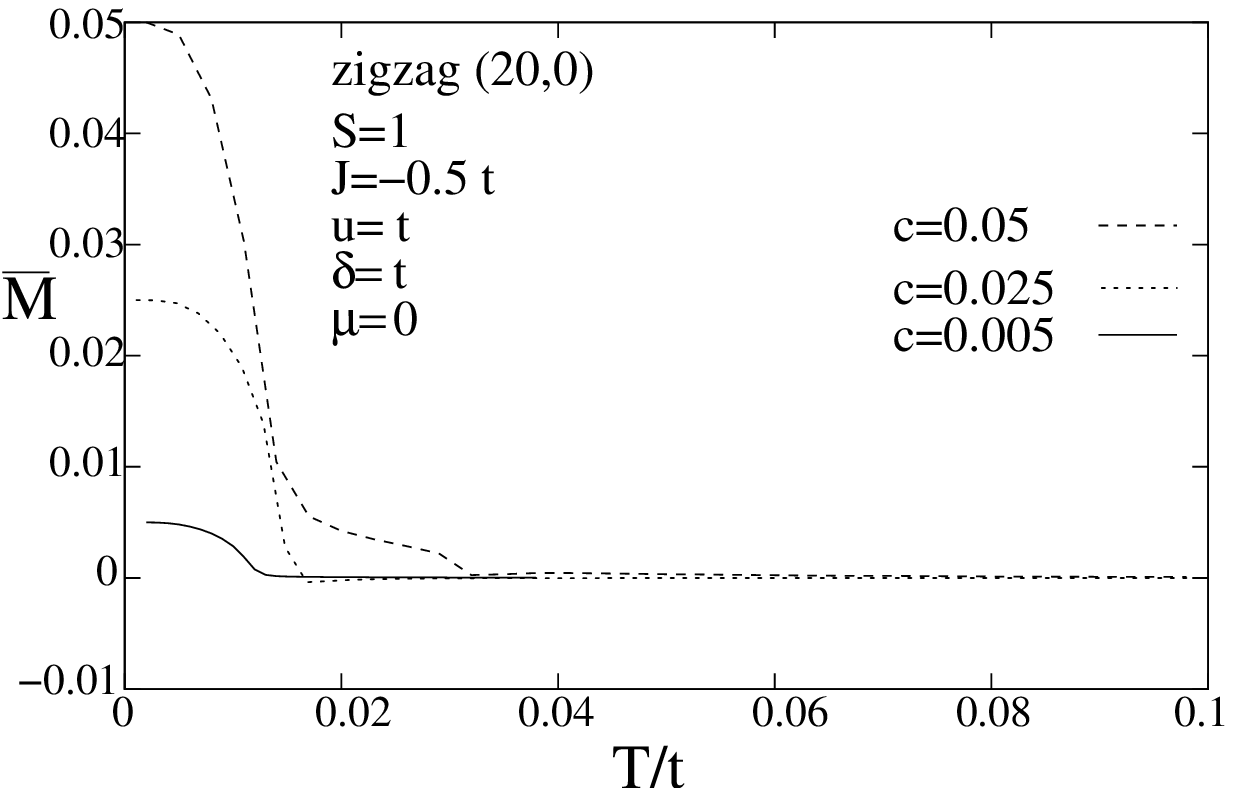  ,width=7.5cm,angle=0}}
\caption{ Show average magnetization of a (10,0) and a (20,0) zigzag SWCNT in terms of temperature at fixed parameters, $S=1$, $J=-0.5 t$, $\mu=0$, $u=t$ and $\delta=t$, for different impurity concentrations, $c=0.005$, $c=0.025$ and $c=0.05$. 
 \label{figure:fig1}}
\end{figure}

\begin{figure}
\centerline{\epsfig{file=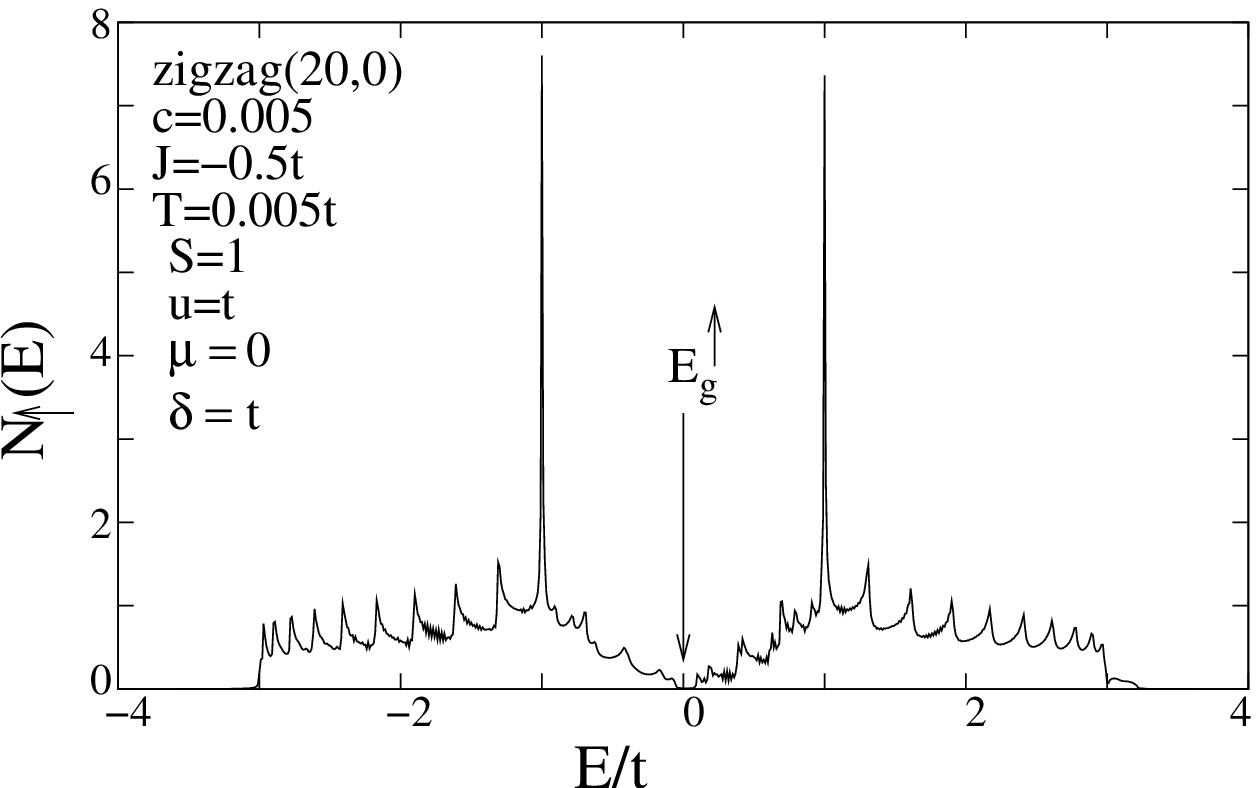  ,width=7.5cm,angle=0}}
\centerline{\epsfig{file=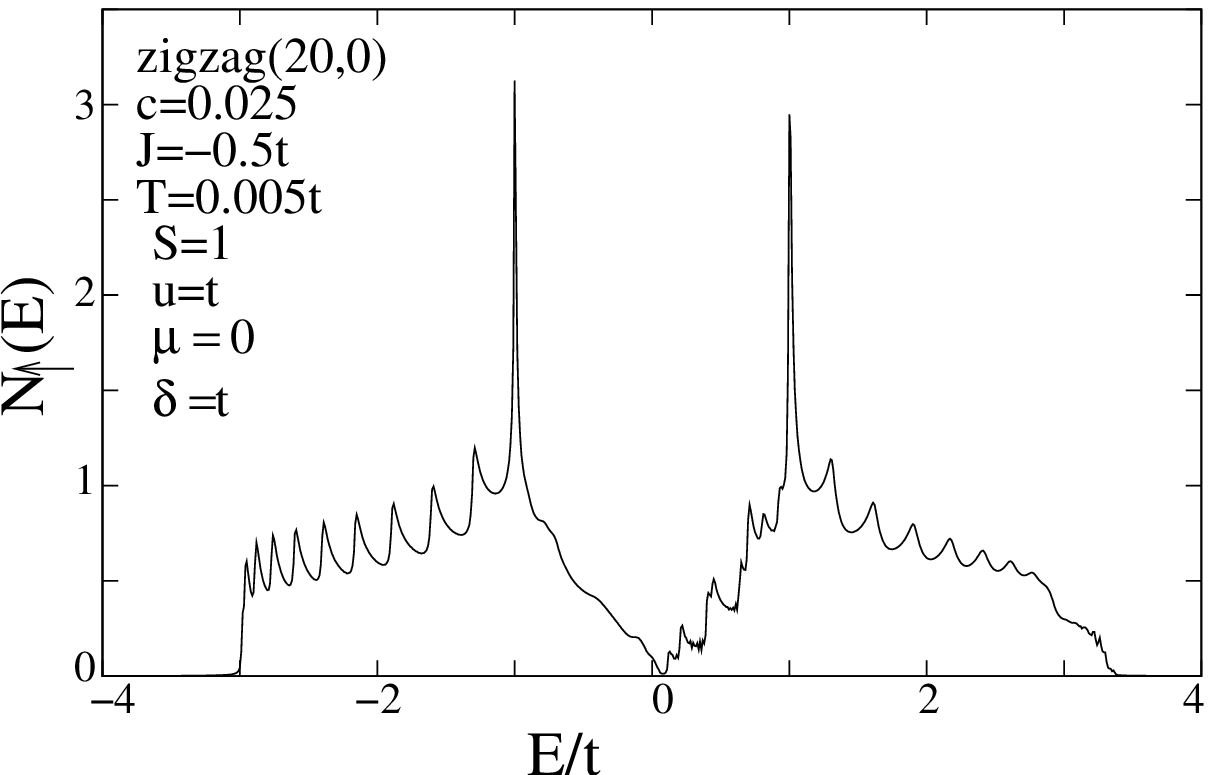 ,width=7.5cm,angle=0}}
\centerline{\epsfig{file=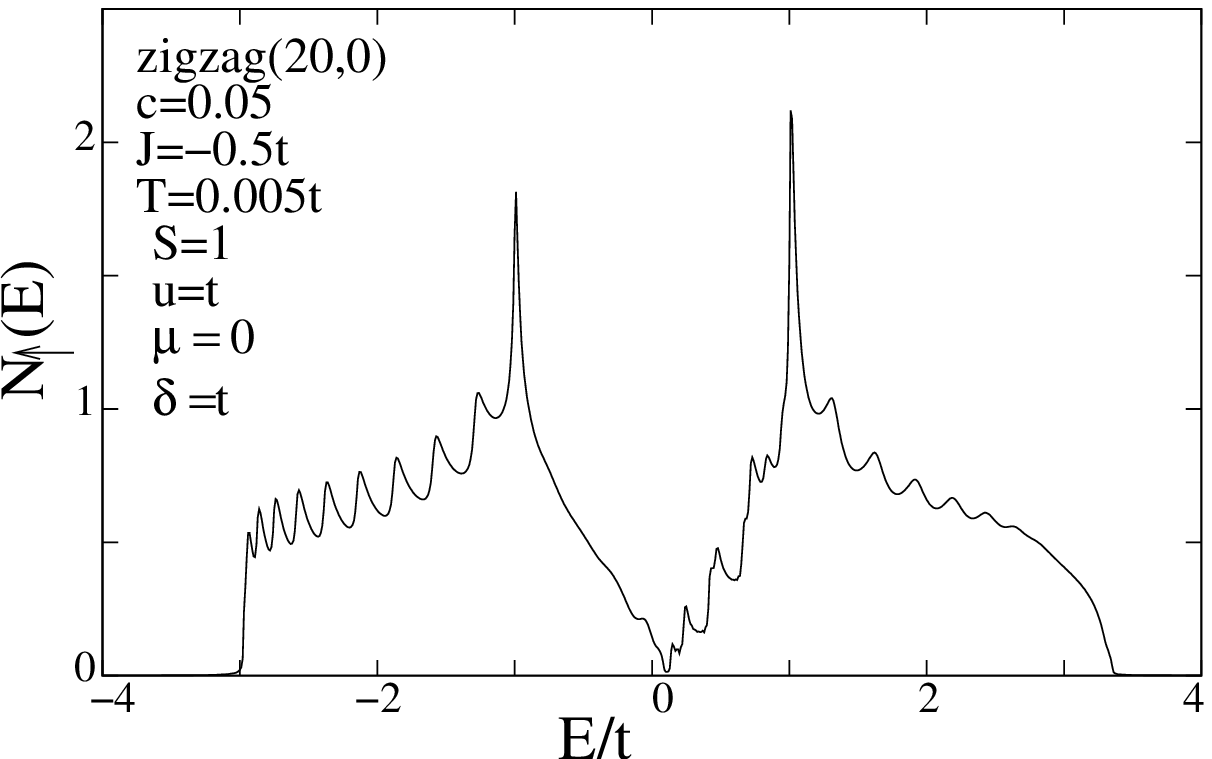 ,width=7.5cm,angle=0}}
\caption{Show density of states for spin up electrons of a (20,0) zigzag SWCNT for different impurity concentrations, $c=0.005$, $c=0.025$ and $c=0.05$ at fixed parameters, $S=1$, $J=0.5t$, $\mu=0$, $u=t$ and $\delta=t$.   
 \label{figure:fig2}}
\end{figure}

\begin{figure}
\centerline{\epsfig{file=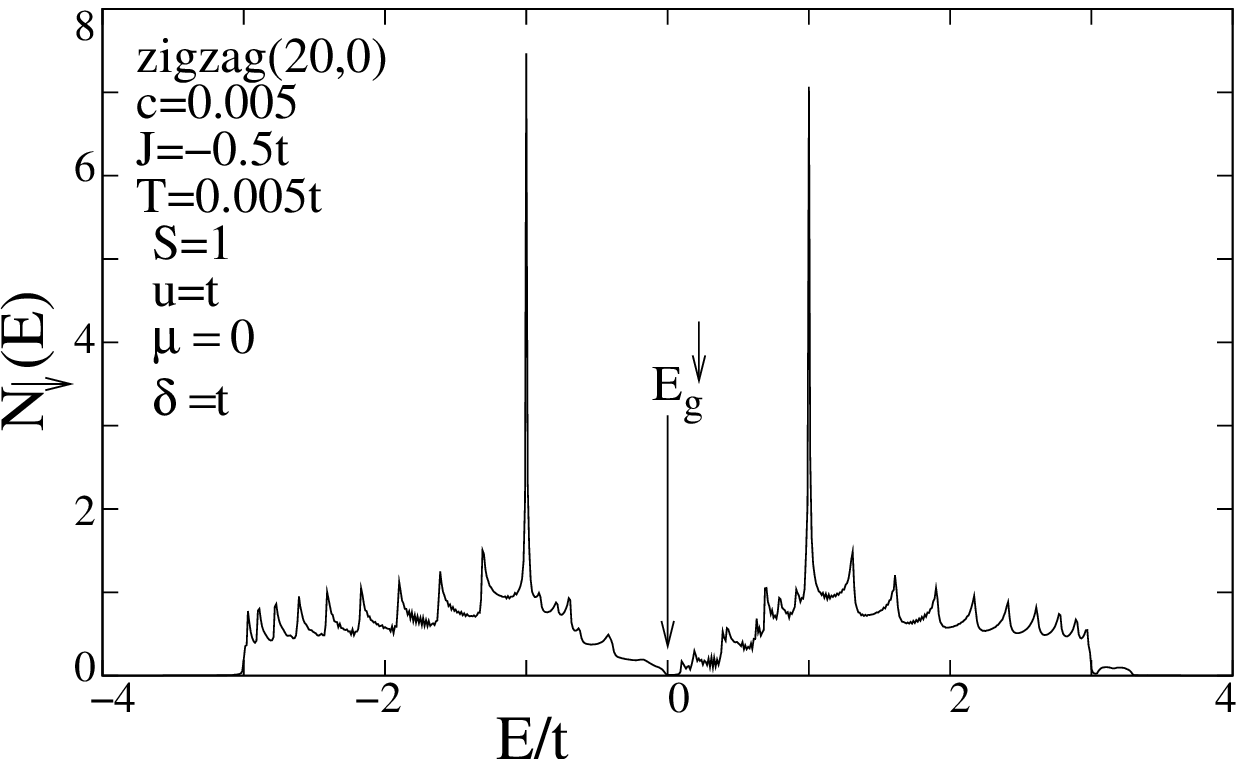  ,width=7.5cm,angle=0}}
\centerline{\epsfig{file=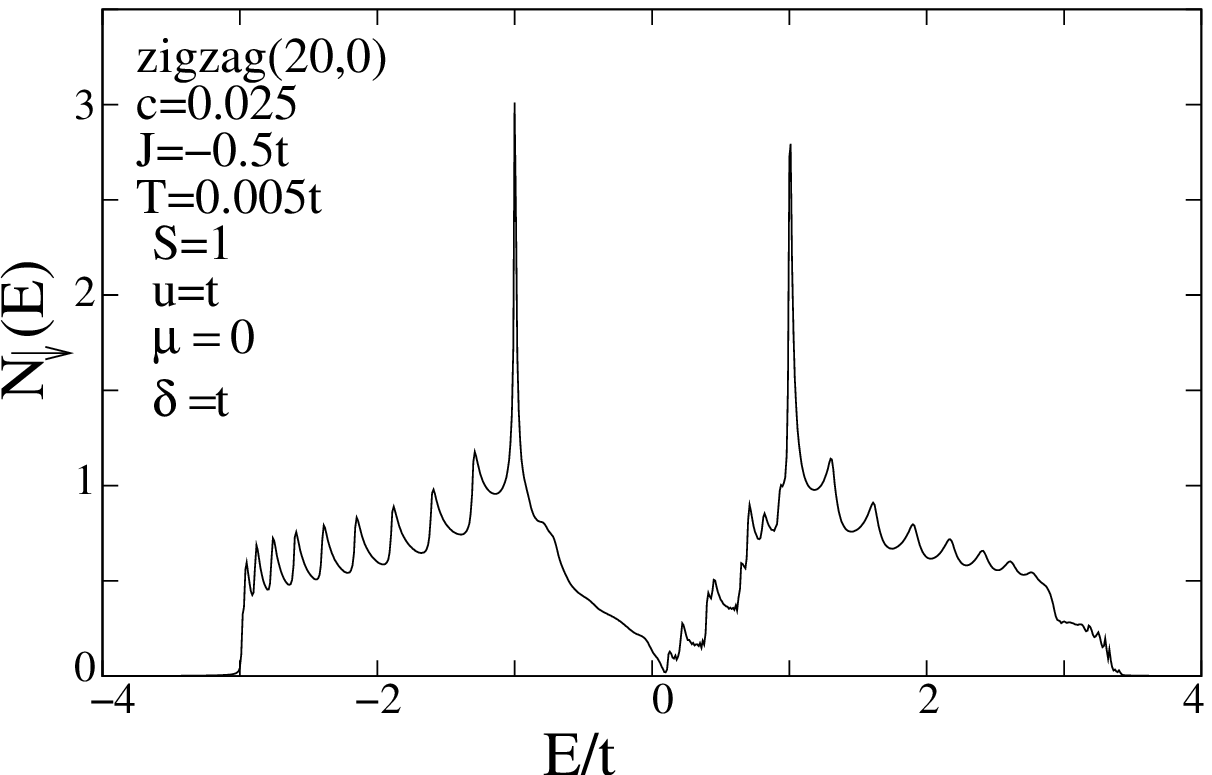 ,width=7.5cm,angle=0}}
\centerline{\epsfig{file=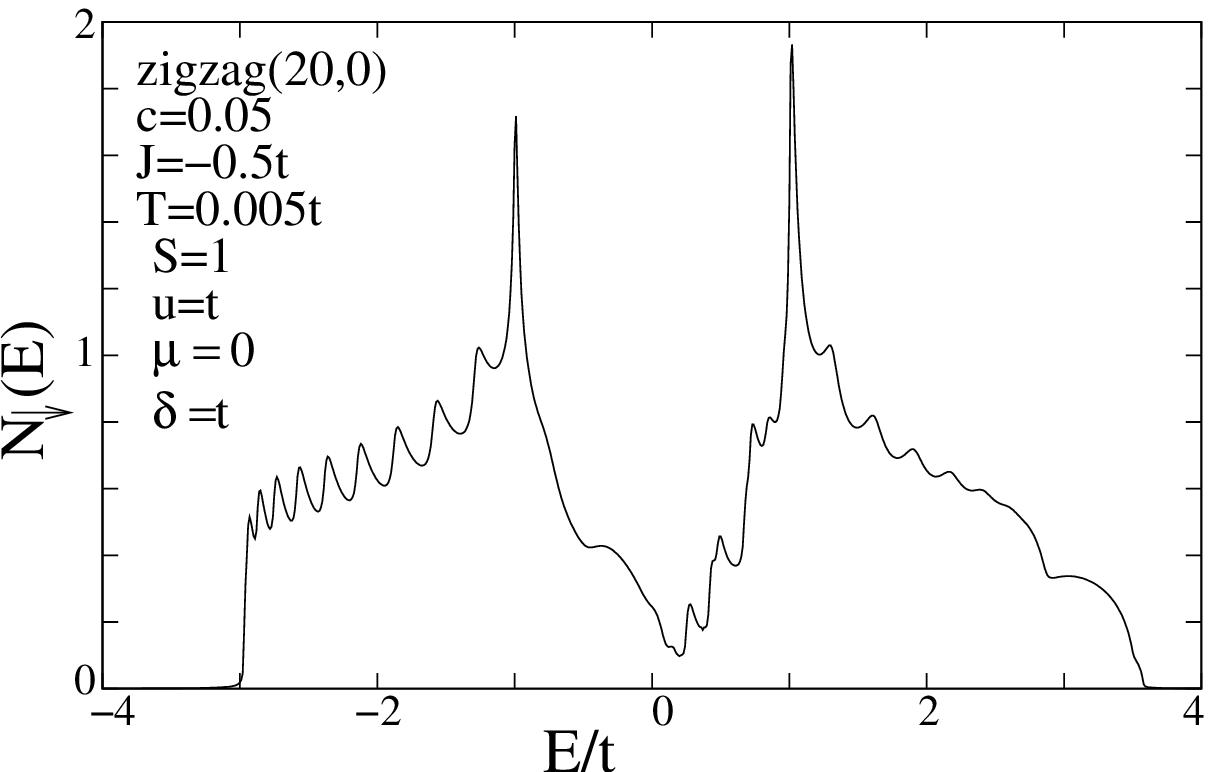 ,width=7.5cm,angle=0}}
\caption{Show density of states for spin down electrons of a (20,0) zigzag SWCNT for different impurity concentrations, $c=0.005$, $c=0.025$ and $c=0.05$ at fixed parameters, $S=1$, $J=0.5t$, $\mu=0$, $u=t$ and $\delta=t$.  
 \label{figure:fig3}}
\end{figure}

\begin{figure}
\centerline{\epsfig{file=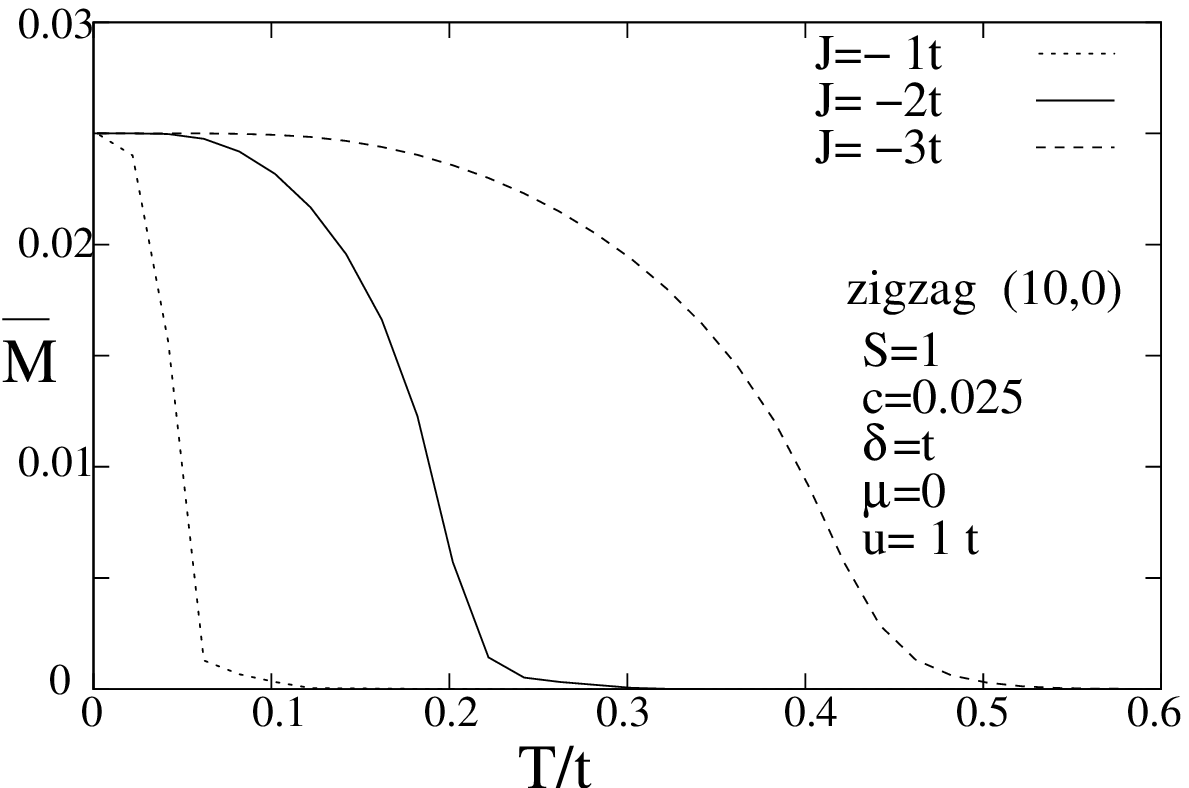  ,width=7.5cm,angle=0}}
\centerline{\epsfig{file=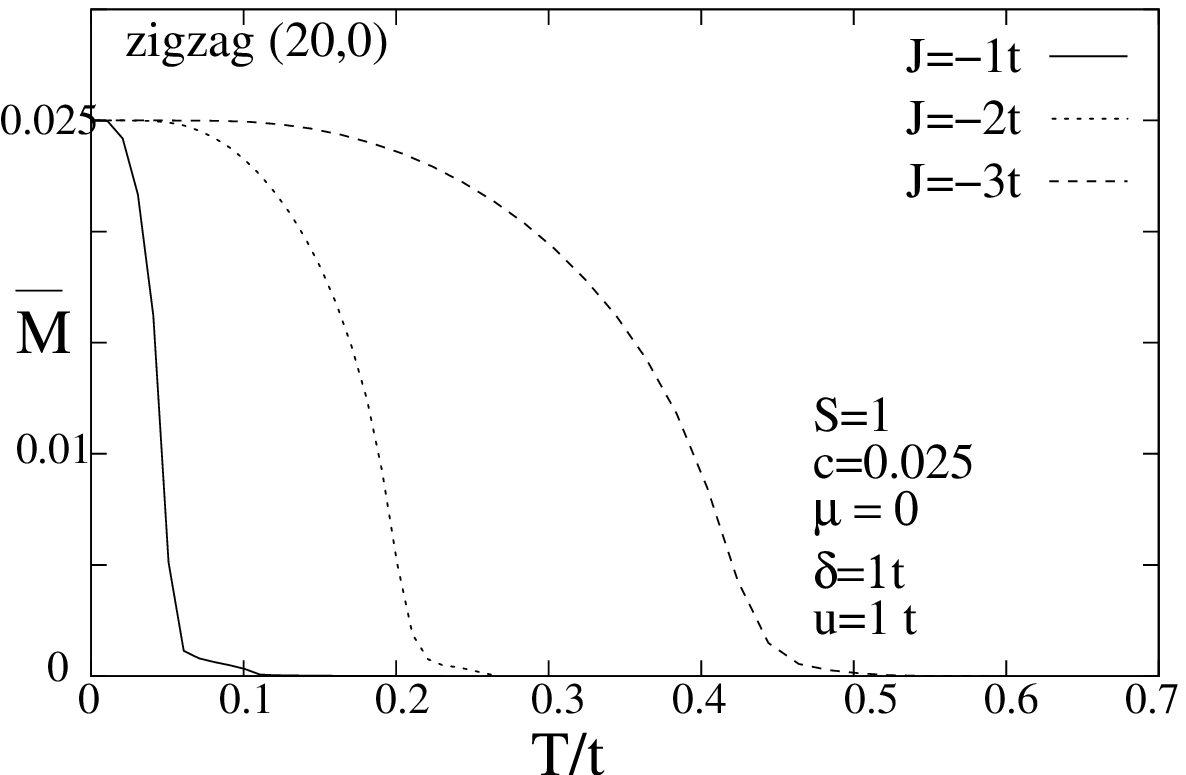  ,width=7.5cm,angle=0}}
\caption{Show magnetization of doped (10,0) and (20,0) zigzag SWCNTs in terms of temperature at fixed parameters, $S=1$, $c=0.025$, $\mu=0$, $u=t$ and $\delta=t$, for different electron impurity spin couplings $J=-t, -2 t$ and $-3t$.  
 \label{figure:fig4}}
\end{figure}

\begin{figure}
\centerline{\epsfig{file=  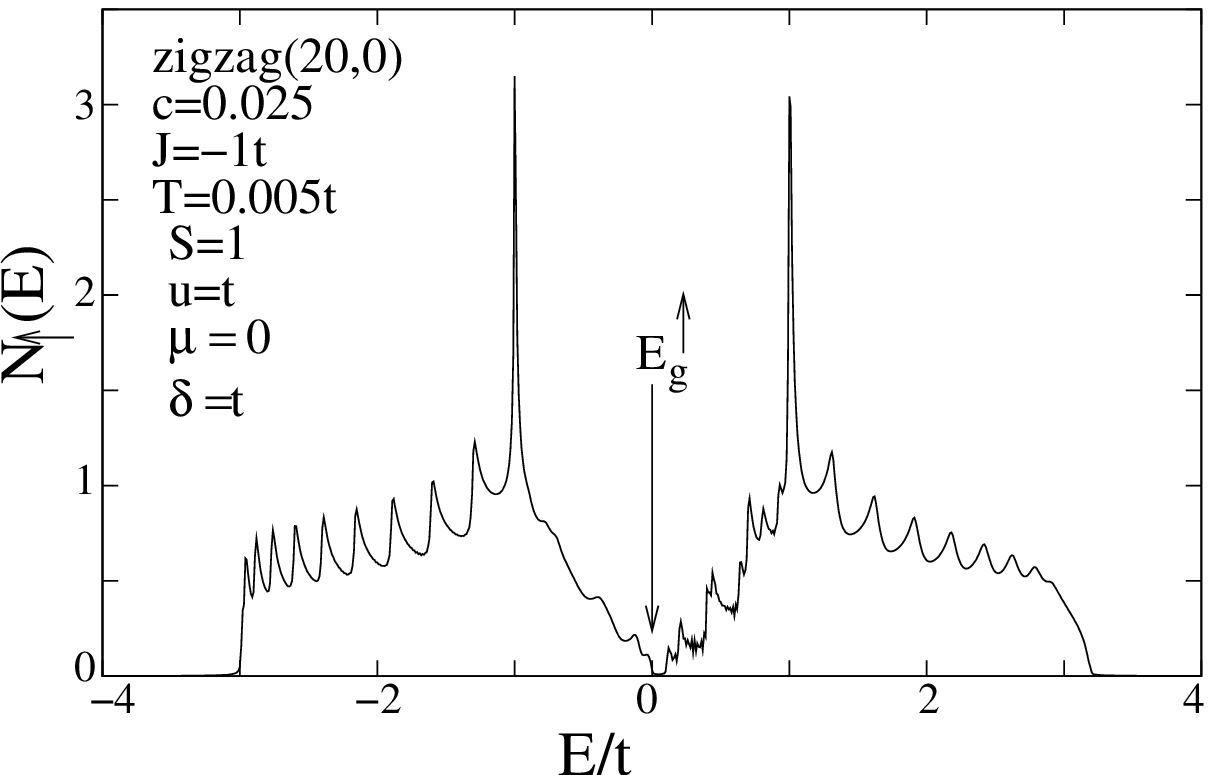 ,width=7.5cm,angle=0}}
\centerline{\epsfig{file=  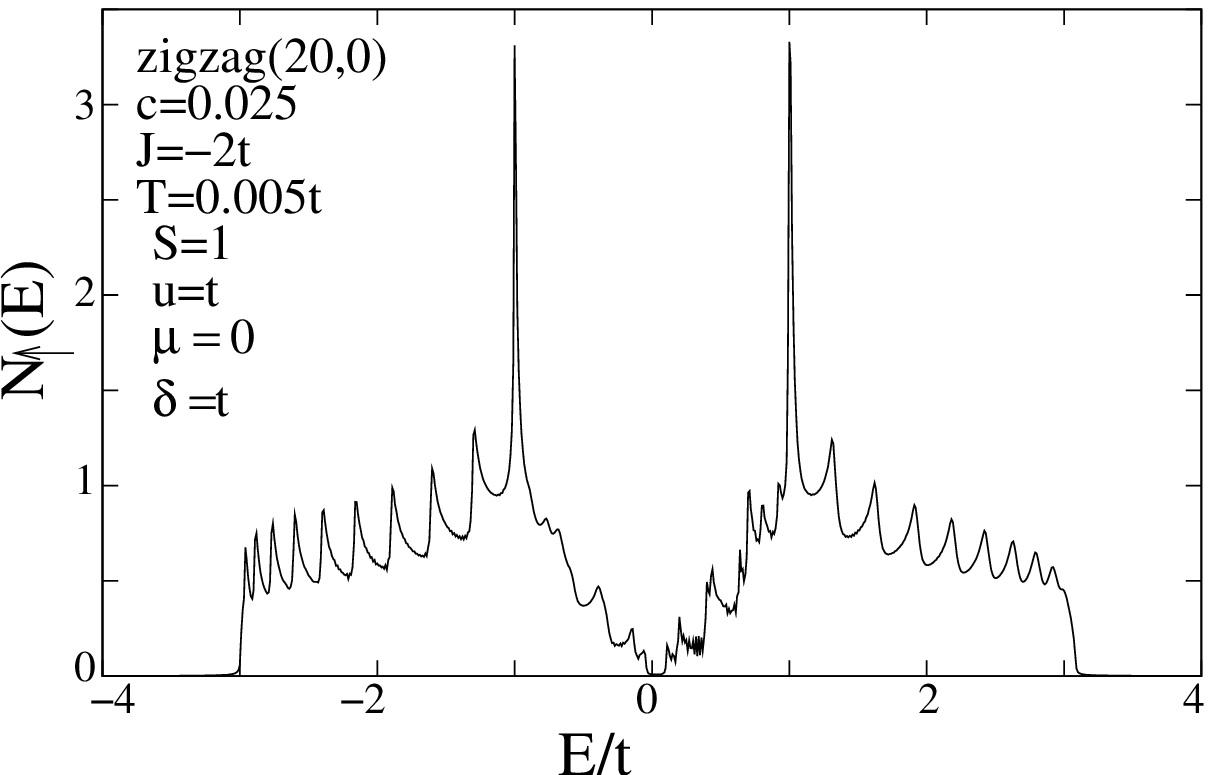 ,width=7.5cm,angle=0}}
\centerline{\epsfig{file=   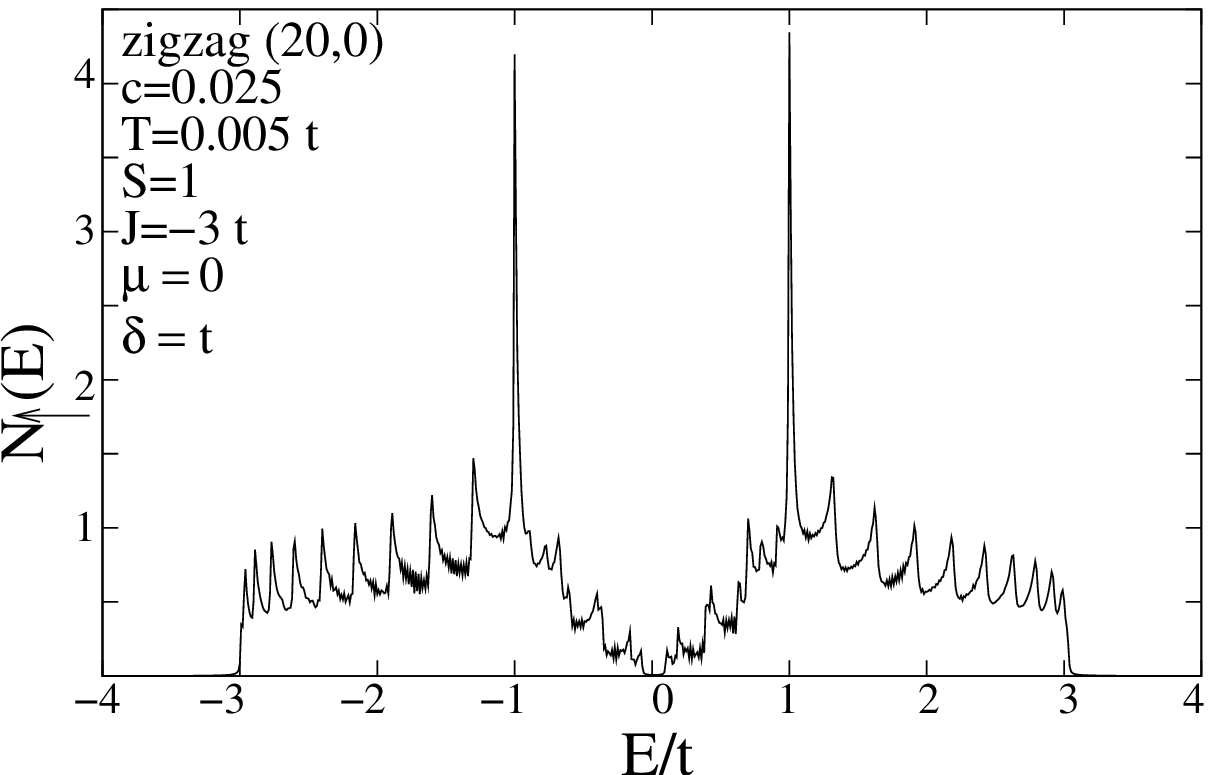  ,width=7.5cm,angle=0}}
\caption{Shows effects of exchange coupling, $J$, on spin up electron density of states of a $(20,0)$ zigzag SWCNT at fixed parameters, $c=0.025$, $S=1$, $u=1$, $\delta=t$ and $\mu=0$. By increasing $J$ the semiconducting gap is increased. 
 \label{figure:fig5}}
\end{figure}

\begin{figure}
\centerline{\epsfig{file= 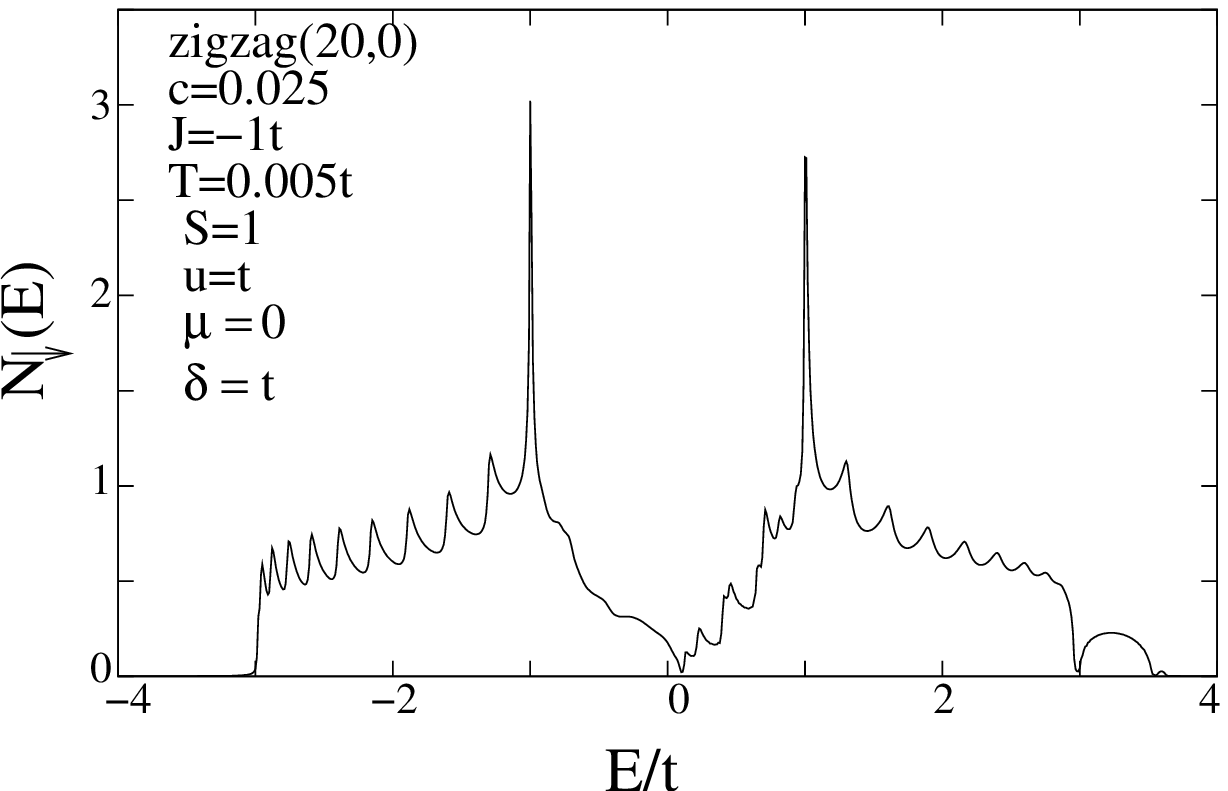 ,width=7.5cm,angle=0}}
\centerline{\epsfig{file= 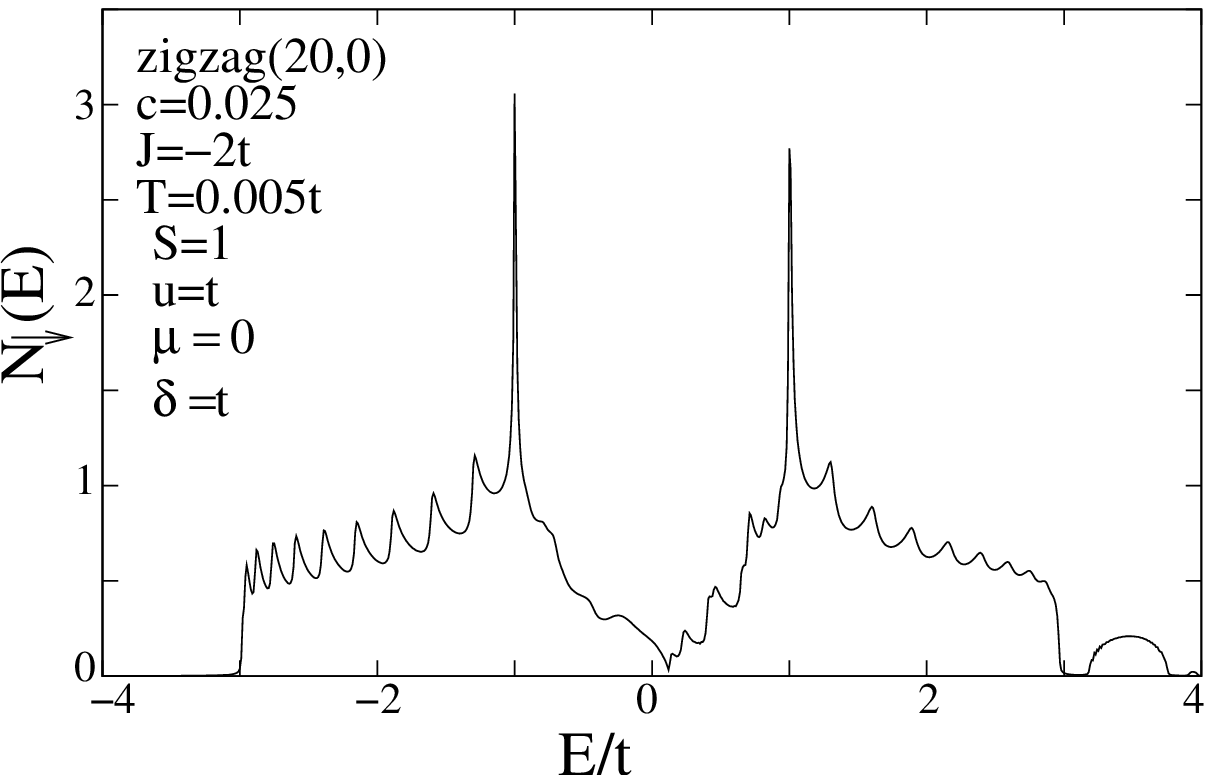 ,width=7.5cm,angle=0}}
\centerline{\epsfig{file=  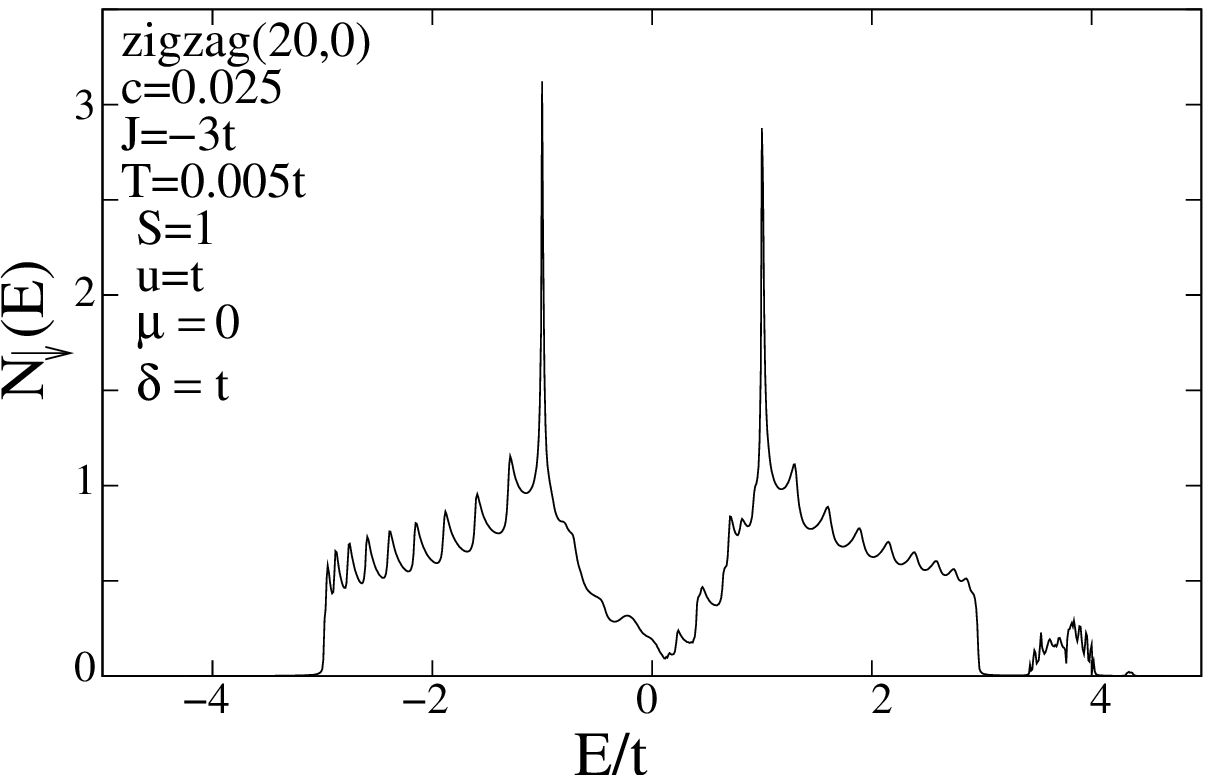 ,width=7.5cm,angle=0}}
\caption{Shows effects of exchange coupling, $J$, on spin down electron density of states of a $(20,0)$ zigzag SWCNT at fixed parameters, $c=0.025$, $S=1$, $u=1$, $\delta=t$ and $\mu=0$. By increasing $J$ the semiconducting gap ,$E^{\uparrow}_{g}$, is completely filled. \label{figure:fig6}}
\end{figure}

\begin{figure}
\centerline{\epsfig{file=  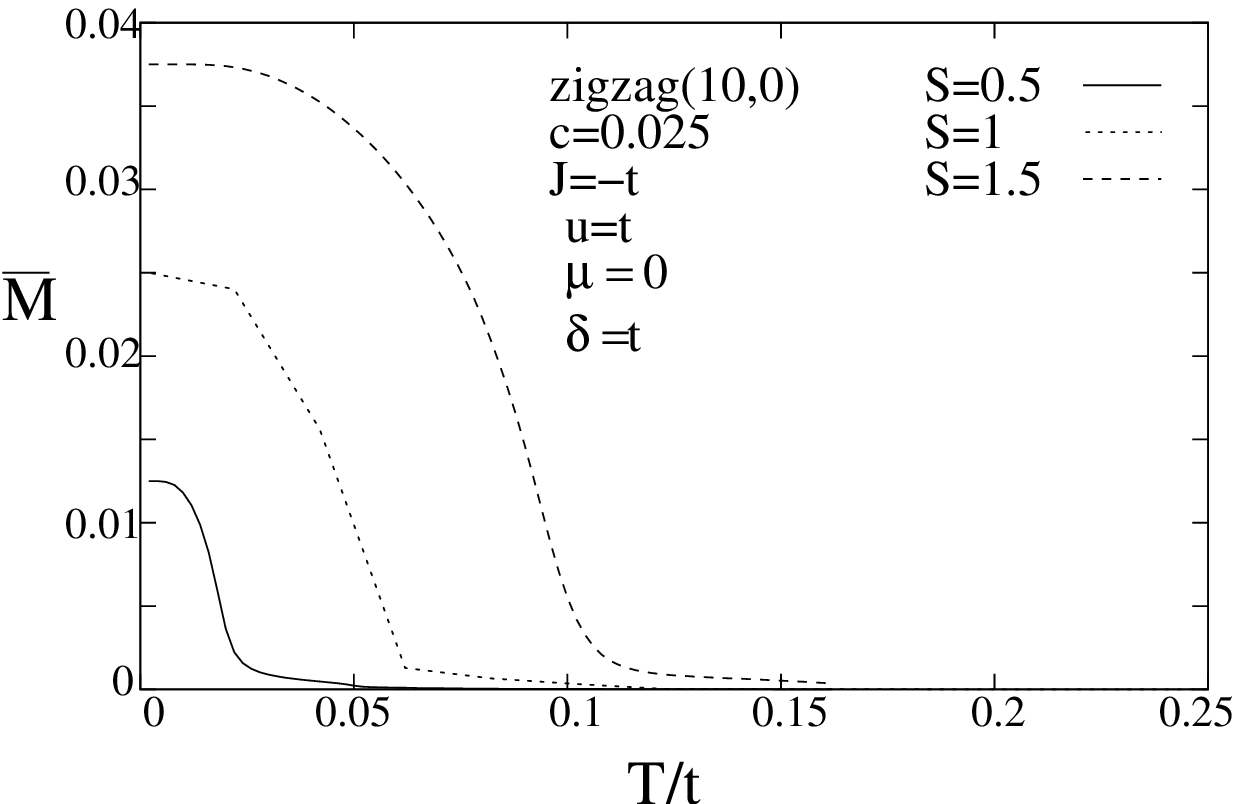 ,width=7.5cm,angle=0}}
\centerline{\epsfig{file=  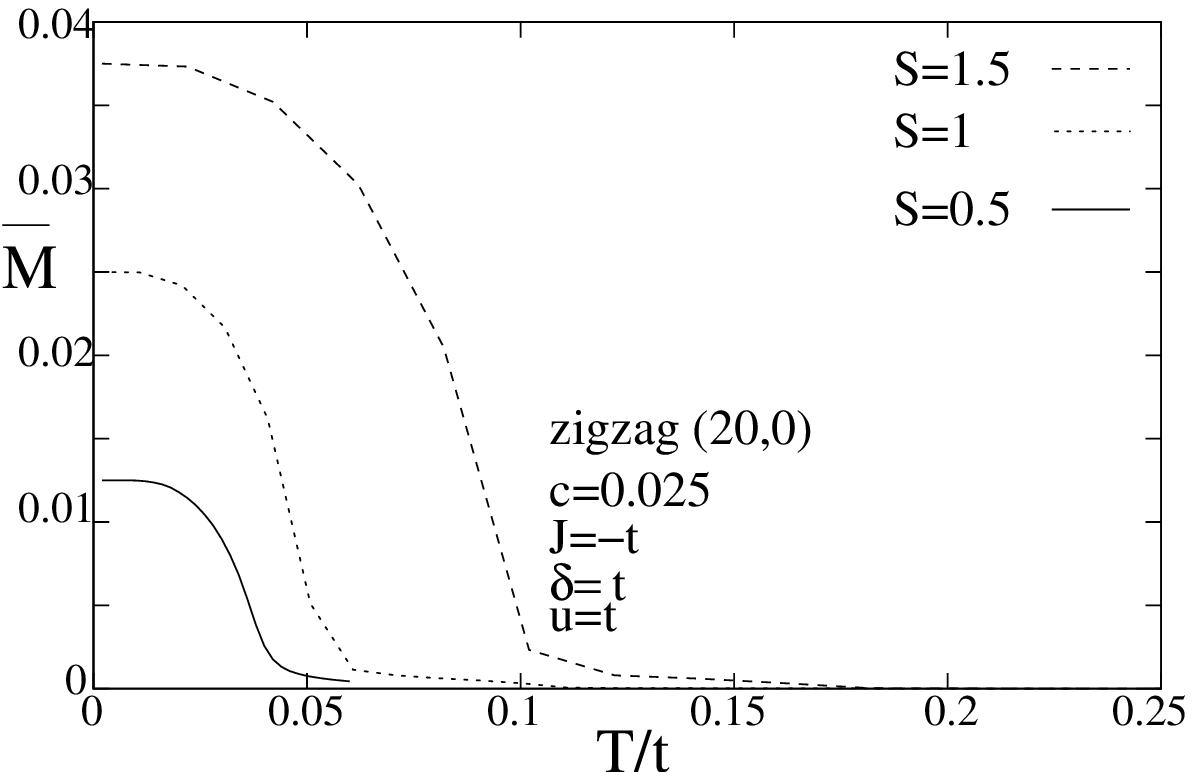 ,width=7.5cm,angle=0}}
\caption{ Show average magnetization of doped (10,0) and (20,0) zigzag SWCNTs in terms of temperature at fixed parameters, $c=0.025$, $J=- t$, $\mu=0$, $u=t$ and $\delta=t$, for different impurity spins, $S=\frac{1}{2}$, $S=1$ and $S=\frac{3}{2}$.
 \label{figure:fig7}}
\end{figure}

\begin{figure}
\centerline{\epsfig{file=  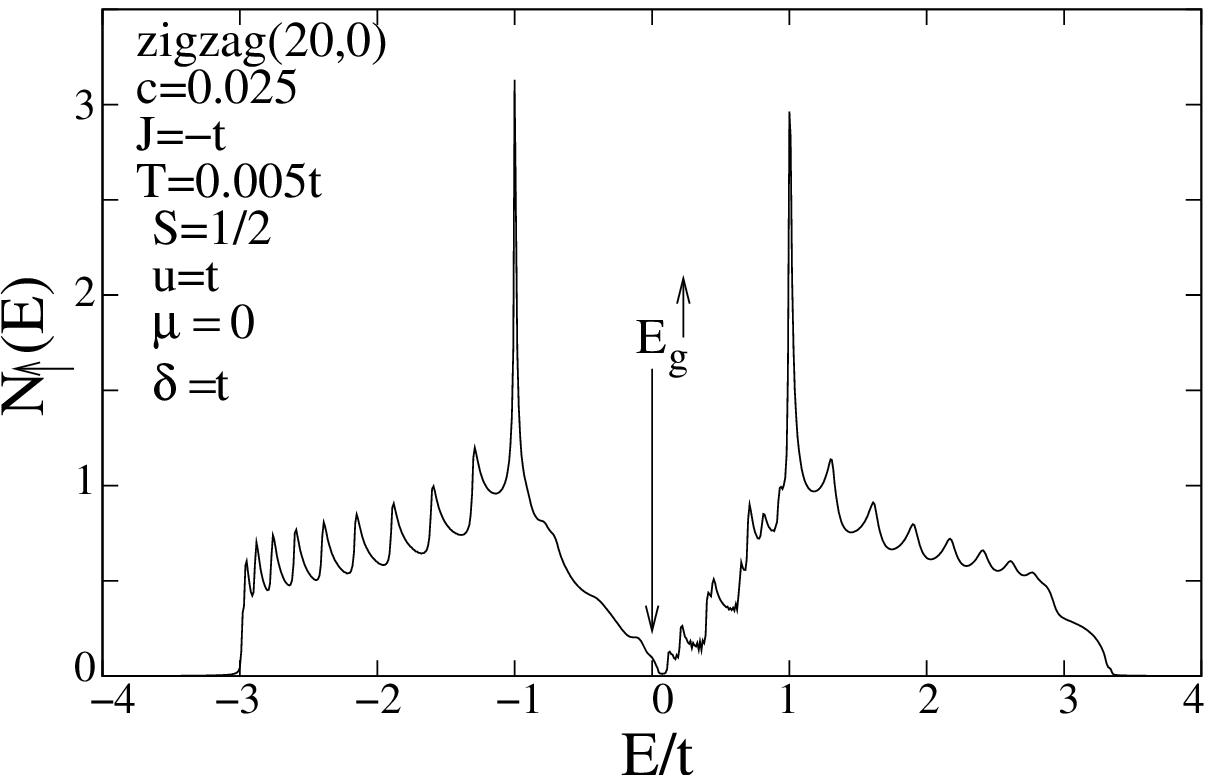 ,width=7.5cm,angle=0}}
\centerline{\epsfig{file=  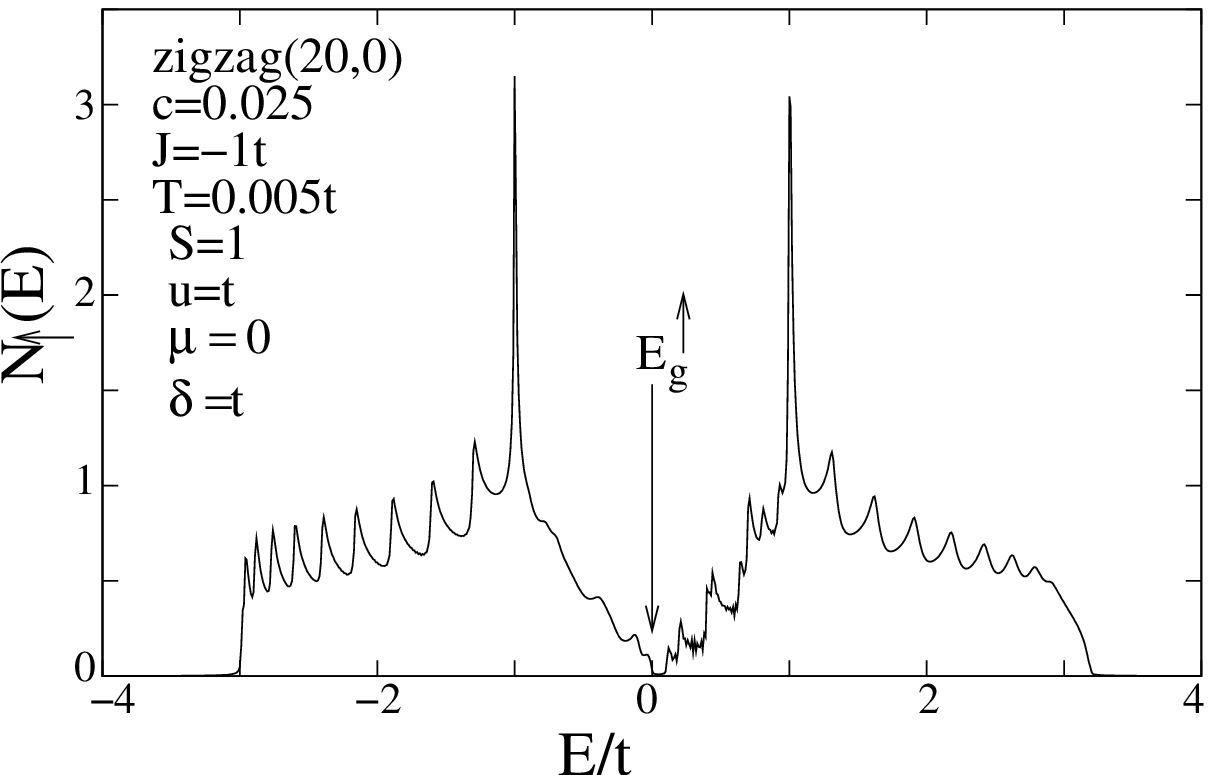 ,width=7.5cm,angle=0}}
\centerline{\epsfig{file= 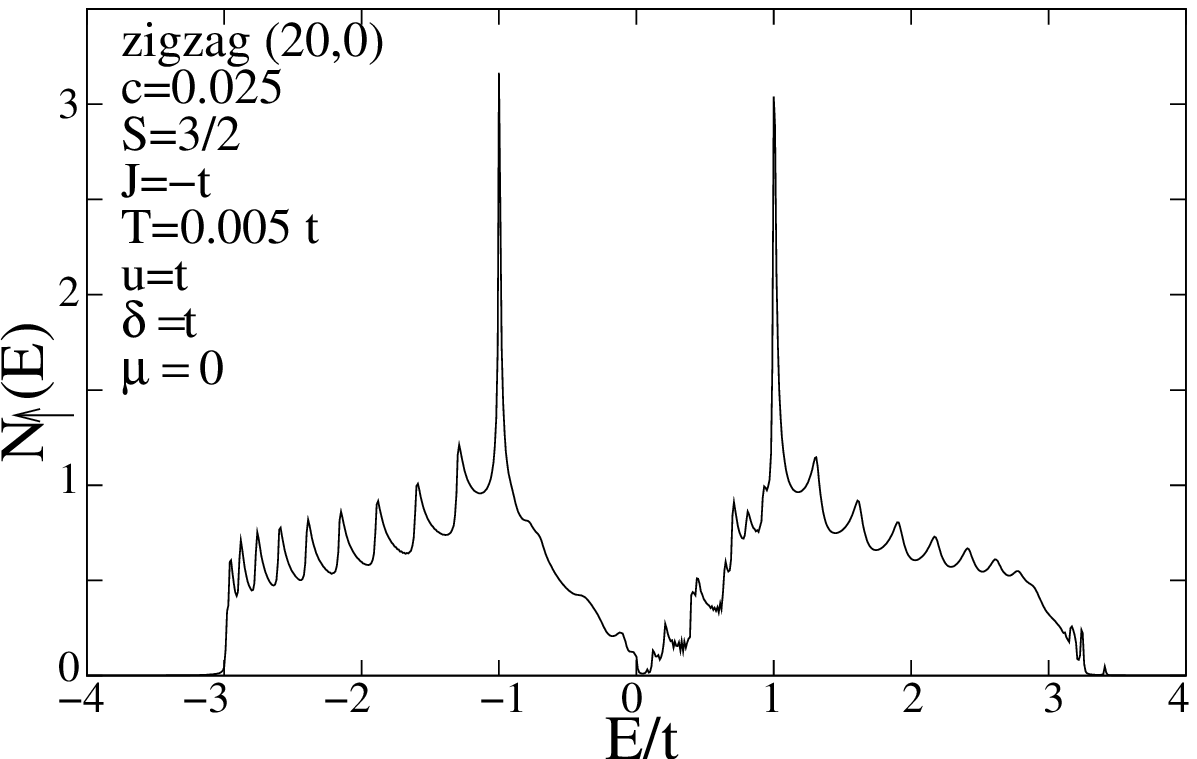  ,width=7.5cm,angle=0}}
\caption{Shows effects of impurity spin magnitude, $S$, on spin up electron density of states of a $(20,0)$ zigzag SWCNT at fixed parameters, $c=0.025$, $J=t$, $u=1$, $\delta=t$ and $\mu=0$. By increasing impurity spin magnitude, $S$, the semiconducting gap in the spin up electrons density of states is increased. This at high, $S$, could lead to a phase transition. 
 \label{figure:fig8}}
\end{figure}

\begin{figure}
\centerline{\epsfig{file=  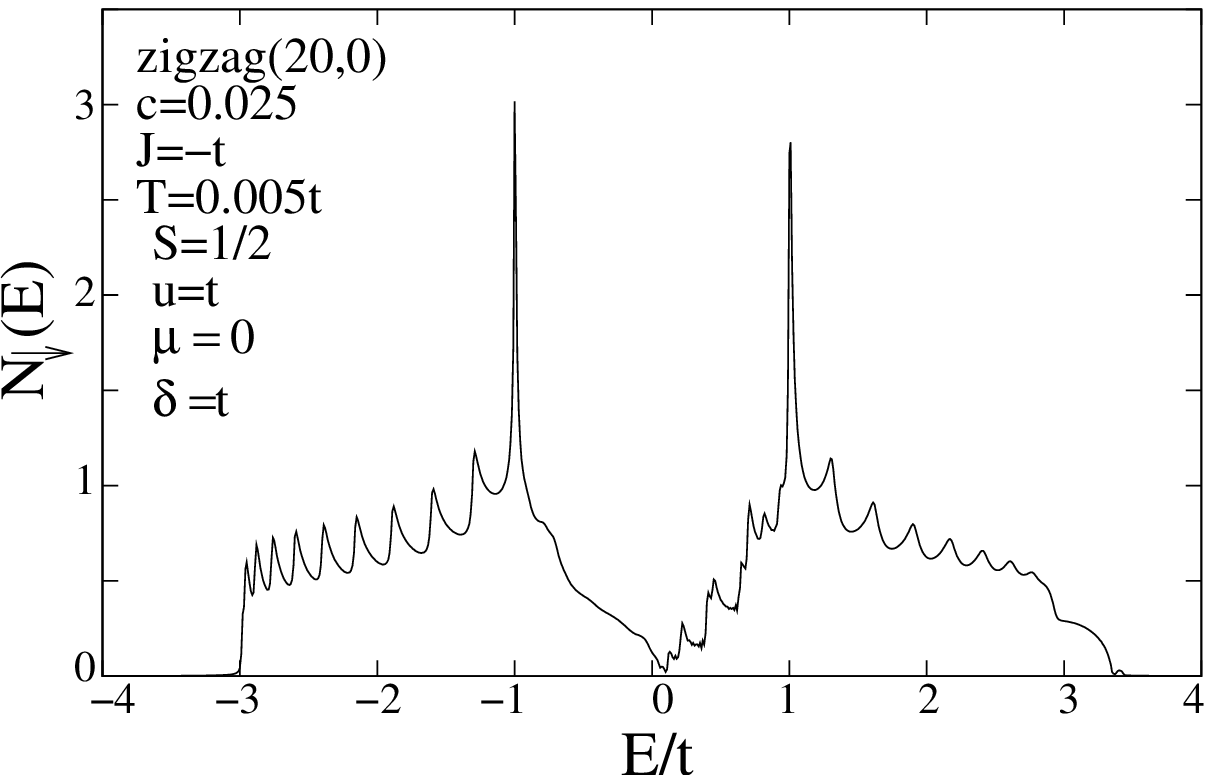 ,width=7.5cm,angle=0}}
\centerline{\epsfig{file=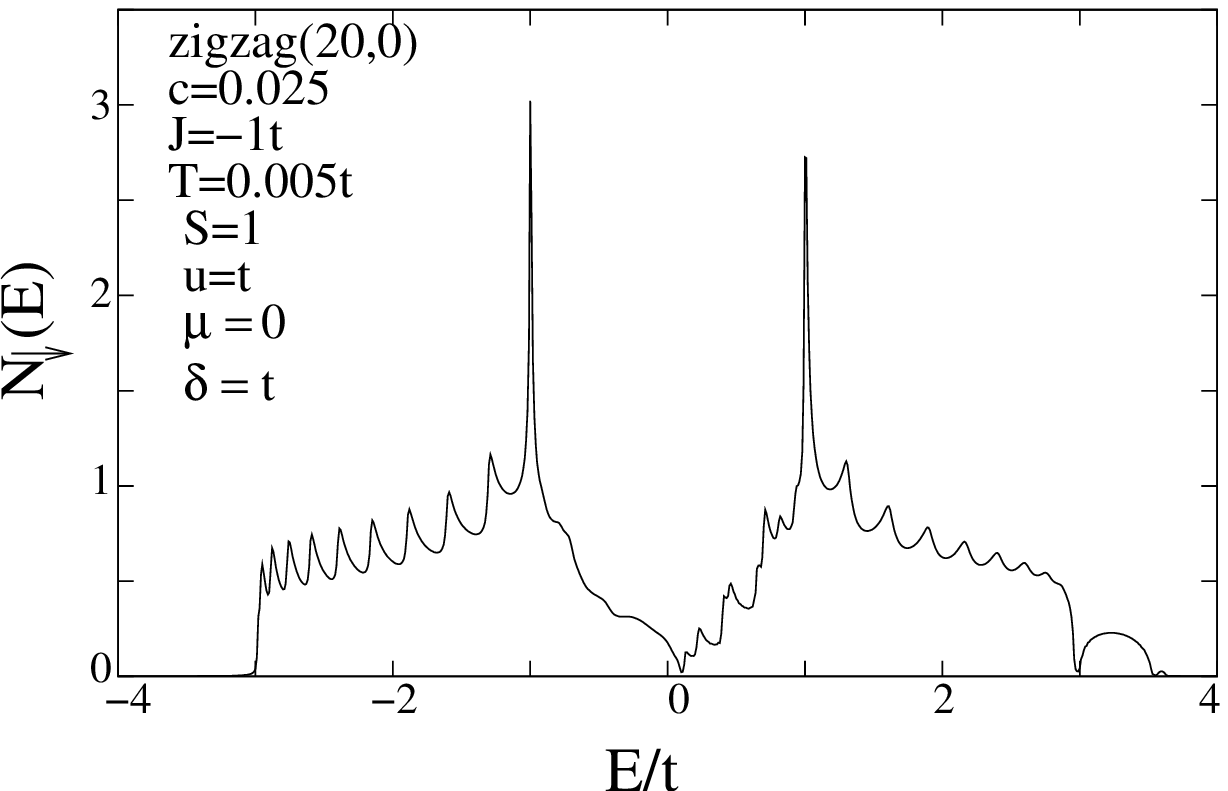 ,width=7.5cm,angle=0}}
\centerline{\epsfig{file= 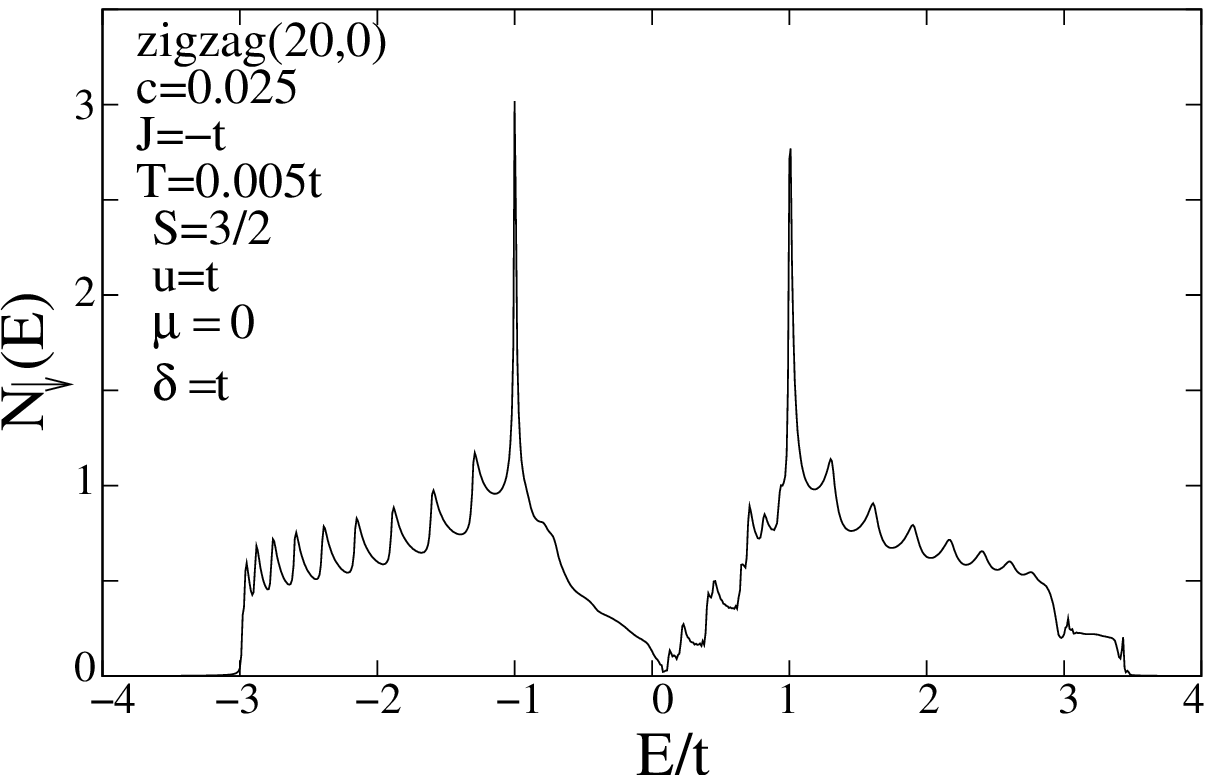 ,width=7.5cm,angle=0}}
\caption{Shows effects of impurity spin magnitude, $S$, on spin down electron density of states of a $(20,0)$ zigzag SWCNT where parameters are fixed at, $c=0.025$, $J=t$, $u=1$, $\delta=t$ and $\mu=0$. By increasing impurity spin magnitude, $S$, the semiconducting gap in the spin down electrons density of states is increased, so the system remain semiconductor. 
 \label{figure:fig9}}
\end{figure}

To investigate magnetic impurity effects on $(10,0)$ and $(20,0)$ zigzag SWCNTs, several cases are considered: first to obtain impurity concentration effects, average magnetization, ${\bar M}$, is calculated in terms of temperature at fixed $J=-0.5 t$, $\delta=t$ and $\mu=0$, Fig.\ref{figure:fig1} illustrate average magnetization in terms of temperature  for different impurity concentrations  $c=0.005$, $c=0.025$ and $c=0.05$.
 Also in this case at a fixed temperature, $T=0.005 t$, spin up and spin down electrons density of states are compared for concentrations $c=0.005$, $c=0.025$ and $c=0.05$. Figs.\ref{figure:fig2} and \ref{figure:fig3} illustrate a semiconductor to metallic phase transition in the spin up electrons density of states due to increasing impurity concentration for the weak exchange coupling $J=-0.5 t$. 
In the case of weak exchange coupling we found by increasing impurity concentration the semiconducting energy gap, $E^{\uparrow}_{g}$, between valance band and conduction band is decreased and a semiconductor to metallic phase transition is occurred. While for strong coupling the spin up density of states is a semiconductor and for spin down is a metal. Second, at a fixed impurity concentration, $c=0.025$, for variable exchange coupling, $J=-t$, $J=-2 t$ and $J=-3 t$, average magnetization is calculated in terms of temperature. Fig.\ref{figure:fig4} shows effect of exchange coupling on average magnetization.
We found for strong impurity couplings, the spin up electrons density of states remain a semiconductor with a small gap, $E^{\uparrow}_{g}$, between valance and conduction bands, value of this gap is increased for higher exchange couplings , while the spin down electrons density of states is continues (it has a metallic behavior). Fig.\ref{figure:fig5} and \ref{figure:fig6} illustrate our result for spin up and down electrons density of states for variable exchange coupling, $J$, and fixed other parameter at $c=0.025$, $S=1$, $\mu=0$, $\delta=t$.    
Third, to observe effects of impurity spin magnitude on our system we fixed, impurity concentration at, $c=0.025$, exchange coupling at $J=-t$, on-site energy at $\delta=t$ and $\mu=0$. Magnetization is calculated in terms of temperature for different impurity spin magnitudes $S=1/2$, $S=1$ and $S=3/2$. Fig.\ref{figure:fig7} show average magnetization in terms of temperature for different spins $S=1/2$, $S=1$ and $S=3/2$. We found by increasing impurity spin value, average magnetization is increased but the spin up electrons density of states has a gap, $E^{\uparrow}_{g}$, which is groves by increasing impurity spin magnitude. Figs.\ref{figure:fig8} and \ref{figure:fig9} illustrate density of states for spin up and spin down electrons in this case. 
\section{conclusion}
We have investigated possibility of making a ferromagnetic semiconductor zigzag SWCNT, by doping magnetic impurities such as Fe, Co, Ni, Mn. The doped $(20,0)$ and $(10,0)$ zigzag SWCNTs are treated in the coherent potential approximation (CPA). To find out role of impurity on our systems three cases are considered, first at fixed, impurity spin magnitude, $S=1$, exchange coupling at $J=-0.5 t$ and temperature $T=0.005 t$ the impurity concentration are chosen variable.  We found by increasing impurity concentration the semiconducting gap, $E^{\uparrow}_{g}$ and $E^{\downarrow}_{g}$, are closed, hence a semiconductor to metal phase transition is take placed. Second, we fixed impurity concentration at $c=0.025$, spin magnitude at $S=1$ and temperature at $T=0.005 t$, then varying exchange coupling,$ J$, we found by increasing $J$ the semiconductor gap, $E^{\uparrow}_{g}$, in the spin up density of states is increased, while gap ,$E^{\downarrow}_{g}$, in the spin down density of states disappeared and have a metallic behavior. Third, at fixed impurity concentration c=0.025, J=-t and T=0.005t, the magnitude of impurity spin, S, varied, in this case we found by increasing spin magnitude , S, the gap in the spin up density of state is increased. In summery we found by magnetic impurity doping on a zigzag SWCNT, one could make a ferromagnetic semiconductor.     

\acknowledgements{This work is supported by the Nano committee of Ministry of Science Research and Technology of Iran.}

\end{document}